# Reduction of radiative lifetime and slow-timescale spectral diffusion in InGaN polarized single-photon sources

Tong Wang,[1,a)] Tongtong Zhu,[2,a)] Tim J. Puchtler,[1] Claudius C. Kocher[1], Helen P. Springbett,[2] John C. Jarman[2], Luke P. Nuttall[1], Rachel A. Oliver[2], and Robert A. Taylor[1]

[1]*Department of Physics, University of Oxford, Parks Road, Oxford, OX1 3PU, UK*

[2]*Department of Materials Science and Metallurgy, University of Cambridge, 27 Charles Babbage Road, Cambridge*

*CB3 0FS, UK*

Non-polar (11-20) *a*-plane quantum dots (QDs) are strong candidates for both > 200 K on-chip ultrafast polarized single-photon generation and the investigation of high temperature semiconductor QD photophysics. In this work, we report progress in the growth of *a*-plane InGaN QDs with a quasi-two-temperature method, which produces smooth epilayers and significantly reduced carrier trapping sites in the proximity of the QDs. Optical characterization has confirmed the ability of such QDs to emit polarized single photons and we have recorded a ~ 45% shorter average radiative lifetime and 65% reduction in the slow-timescale spectral diffusion compared to previous QDs. This growth method is an important development of the non-polar *a*-plane InGaN platform, opening up more possibilities in single-photon, lasing, and fundamental investigations.

## I. INTRODUCTION

Rapid and significant progress has been made in the fundamental understanding and development of polarized single-photon sources [1], which are indispensable elements in the science of quantum information processing [2–4] and optical quantum computing [5,6]. While developments in systems such as material defects [7–11], carbon nanotubes [12], single molecules [13], and perovskites [14] provide several alternative visions of single-photon applications, semiconductor quantum dots (QDs) [15–17] remain the most practical and attractive solution owing to their much more established growth routines and straightforward integration in solid-state platforms. In particular, III-nitride QDs [18,19] remove the restriction of semiconductor single-photon operation at cryogenic temperatures. The ternary alloy InGaN is needed [20,21] in order to achieve flexibility in the emission energies in the visible range (whereas pure GaN QDs and indeed AlGaN QDs typically emit in the ultra-violet), for highly efficient short-range transmission and detection. With the further development of non-polar (11-20) InGaN QDs [22], intrinsic optical polarization control [23,24] and minimization of the undesirable internal

---

a) T.W. and T.Z. are co-first authors and contributed equally to this work. Electronic mails: tong.wang@physics.ox.ac.uk; tz234@cam.ac.uk.

fields have been achieved [25]. The achievable polarization control in the non-polar system lifts the 50% solid-state QD external quantum efficiency limited imposed by external polarizers, whilst reduced the radiative lifetime, allowing for GHz repetition rates. All this is achievable at temperatures above the on-chip Peltier cooling threshold of 200 K [26]. From a fundamental perspective, the non-polar InGaN system also opens up the underexplored field of high temperature QD photophysics, such as the recent observation of temperature-dependent fine structure splitting [27]. Therefore, both an in-depth understanding and continued development of the system are essential.

Long radiative lifetimes (> 1 ns) [28] and significant slow-timescale spectral diffusion (100 to 500 μeV) [29] are two of the factors that have adversely affected the development of polar *c*-plane InGaN QDs. Several approaches, such as the use of non-polar *a*- and *m*-plane InGaN [25,30], have alleviated these problems to different extents. However, no InGaN systems have achieved sub-200 ps radiative lifetime or sub-20 μeV spectral diffusion [31] as observed in separate GaN QD systems. The incorporation of indium in the InGaN alloy requires growth temperatures lower than optimal, introducing more defects and carrier trapping sites that adversely affected the optical properties of the material [32,33]. This problem is especially pronounced in the growth of *a*-plane InGaN QDs. The previously developed modified droplet epitaxy (MDE) [22] growth method not only requires temperatures much lower than those typically used for polar *c*-plane InGaN [34,35], but also predicates on an annealing procedure designed to induce InGaN epilayer decomposition into gallium-indium droplets (a disruption to epilayer morphology which inevitably creates more carrier traps) enabling subsequent dot formation upon capping.

In this letter, we report significant progress in realizing *a*-plane InGaN QD fabrication with considerably reduced carrier trapping sites using a quasi-two-temperature (Q2T) method, and contrast the differences in the optical properties of these polarized single-photon emitters with those grown by MDE. Q2T QDs have exhibited slow-timescale spectral diffusion as low as 17 μeV, with an average below 50 μeV, as well as average radiative lifetime below 200 ps. These are significant improvements on *a*-plane InGaN QDs grown by the MDE route and are comparable to GaN systems.

## II. EXPERIMENT

Both MDE and Q2T growth methods use metal-organic vapor phase epitaxy, and the sequences are similar until the growth of the InGaN epilayer. With trimethylgallium, trimethylindium and ammonia as precursor gases, a 16-ML (Q2T) or 10-ML (MDE) InGaN epilayer is grown at low temperatures of 690 and 675 °C, respectively. Immediately afterwards, Q2T samples are capped by an initial 2-nm GaN layer at the InGaN growth temperature, after which the temperature is ramped and stabilized at 860 °C in 90 s before the final capping with 8-nm GaN. This is in contrast to the MDE method, in which a 30-s annealing in $N_2$ atmosphere immediately follows the InGaN epilayer growth, before growth of an initial 10-nm GaN capping layer for dot formation at the same temperature and a final 10-nm capping layer grown at 1050 °C. Based on atomic force microscopy (AFM) images of uncapped QD samples shown in Figure 1(a), the MDE QDs have an average diameter of 35 nm and height of 7 nm on top of an underlying interlinking network of InGaN nanostructures – the aforementioned morphology disruptions – along approximately the [0001] direction due to the annealing and decomposition processes. The Q2T QDs exhibit a bimodal QD size distribution (heights > 10 nm or < 2 nm), as shown in the statistics in Figure 1(a) and 1(b), on a much smoother epilayer. To isolate single QDs and increase the light extraction efficiency for µPL experiments, identical nanopillars structures (~180 nm in diameter) were made by silica nanosphere masks and inductively-coupled plasma reactive-ion etching to a depth ~ 350 nm, as shown by the scanning electron microscope (SEM) images in Figure 1(c) [23]. The resultant Q2T QDs pillars have a better material quality and fewer surrounding carrier trapping sites overall due to (i) much smoother epilayer morphology, (ii) less initial indium incorporation (690 vs. 675 °C, and noting that non-polar InGaN is extremely sensitive to growth temperatures [33]), and (iii) higher (860 vs. 690 °C) dot formation temperature.

Despite the variations in QD sizes, we have previously established several optical properties of non-polar InGaN QDs, such as radiative lifetime [25] and optical polarization [23,24], are highly insensitive to these differences. However, the creation of carrier traps in the proximity of these QDs does affect their emission characteristics. In Figure 1(d), we illustrate the cases in which there are no (ideal, left) and some (right) external carriers around an ideal *a*-plane InGaN QD. The field created by the carriers would cause a slight shift in the energy profile, resulting in an instantaneous reduction of both the spatial overlap of the electron and hole wavefunctions and the emission energy. Hence, a shorter radiative recombination lifetime and less significant spectral diffusion would be expected from a QD with fewer carriers around it, i.e. Q2T QDs. An illustration of the polar InGaN case is shown in Figure 1(e) so as to highlight the differences in the changes in their energy profile and their much more drastic extent, due to opposite strain direction at the material interface and their resultant strong piezoelectric fields.

To assess and compare the optical properties of the Q2T and MDE QDs, we have performed micro-photoluminescence (µ-PL) experiments. The samples were placed in a closed-cycle cryogenic system (AttoDRY800) that maintains a stable temperature of 4.7 K. 1 ps pulses at 800 nm from a 76 MHz repetition-rate Ti:Sapphire laser were used to provide two-photon excitation of the sample, for enhanced relative absorption cross-section of the QDs [36]. The laser beam is focused by a 100× NIR objective (0.5 NA) to a ~ 1 µm spot for sample excitation. A half-wave plate and a fixed polarizer, identical to our previous investigation [26], are used for polarization studies. The sample PL enters a Shamrock 500i spectrograph (0.5 m, 1200 l/mm) and is imaged onto an Andor iDus 420 charge-coupled device, with a pixel resolution for linewidth measurements up to ~300 µeV (wavelength-dependent) accuracy. Alternatively, the PL is detected by a photomultiplier tube (PMT) for time-correlated single-photon counting, or by a standard Hanbury Brown and Twiss (HBT) setup with tunable bandpass filters for photon correlation experiments.

## III. RESULTS AND DISCUSSION

We first confirm the ability of Q2T QDs to act as sources of polarized single photons, akin to their MDE counterparts. Time-integrated and polarization-resolved µ-PL spectra from typical self-assembled Q2T and MDE QDs are shown in Figure 2(a). The studied dots have characteristic sharp emission profiles, which overlap with the QW emission attributed to the underlying InGaN epilayer. As QD formation relies on the underlying QW in both growth routines, emission at any QD wavelength will always contain QW components. For MDE QDs, the intensity of the underlying fragmented QW emission would be lower, due to smaller area of QWs (c.f. Figure 1(a)) captured by the same laser spot. The ratio of the Q2T QD's integrated intensity to the total intensity over the QD spectral window has been estimated to be ~ 67%, lower than the MDE value of ~ 84%. The deterministic polarization axis for *a*-plane QDs has been previously confirmed to be along the crystal *m*-direction with MDE samples [23]. For the studied Q2T QD, the PL also has maximum intensity when the polarizer is parallel to the *m*-axis, and gradually decreases to a minimum as the PL is aligned towards the *c*-direction. An example of the intensity variation with polarizer angle, as shown in the inset of Figure 2(a), has been fitted with Malus' Law, $I(\theta) = I_{max}\cos^2\theta + I_{min}\sin^2\theta$, where $\theta$ is the polarizer angle, $I_{max}$ and $I_{min}$ correspond to the PL intensities parallel and perpendicular to the *m*-axis respectively. According the definition of the degree of optical linear polarization (DOLP), $DOLP = (I_{max} - I_{min})/(I_{max} + I_{min})$, this result directly confirms the emission of polarized photons with a DOLP of 0.92 ± 0.05 and polarization axis along the crystal *m*-direction (0° in the optical setup).

To further ascertain the Q2T polarization properties with statistical significance, 50 Q2T and 50 MDE QDs have been studied individually and their DOLP compared. The Gaussian distributions in Figure 2(b) have mean values of 0.87 ± 0.09 and 0.88 ± 0.08, for MDE and Q2T QDs respectively, which are in close agreement with our previous *a*-plane MDE InGaN QD polarization studies [23]. Furthermore, the widths of the two Gaussian distributions are also similar, indicating that the unavoidable dot size variations and shape anisotropies, especially those have originated from these two different growth routines (c.f. Figure 1(a)), do not affect their polarization properties significantly. The identical polarization degrees, small standard deviation, and fixed polarization axis indicate that QDs grown using the new Q2T method are equally good polarized photon emitters when compared to MDE dots.

In order to provide direct evidence for the emission of single photons in both types of QDs, standard HBT experiments were performed and the events recorded form histograms of photon statistics shown in Figure 2(c). For the QDs in Figure 2(a), uncorrected raw $g^{(2)}(0)$ of 0.62 (Q2T) and 0.39 (MDE) are found. The relatively higher $g^{(2)}(0)$ of the Q2T QD is attributed to the larger background QW emission (c.f. Figure 2(a)) explained earlier. Due to large errors in estimating the dot-to-background ratios, partially owing to the variable wavelength-dependent bandpass filter functions, background corrections were not attempted. In both cases, rapid re-population and emission from QDs also contribute to higher raw $g^{(2)}_{cor}(0)$ values. [18].

We now look at changes in spectral diffusion and radiative lifetime. Firstly, it is important to note that there are two types of spectral diffusion happening on different timescales [29]. One occurs on fast-timescale (ns) and is caused by the presence of mobile carriers in the quantum well and causes broadening of QD exciton transition linewidth to well above their Fourier-limited values [37]. The other involves carriers trapped locally for a much longer time (ms to s), causing slow-timescale spectral diffusion as manifested in random jumps in the QD emission energy. For both Q2T and MDE, the QD emission is reliant on the presence of underlying quantum wells, into which carriers are optically injected (below the GaN bandgap). Hence, there should be minimal differences in their fast-timescale spectral diffusion caused by the growth methods. On the other hand, the size, geometry and indium content fluctuations of the self-assembled quantum emitters would all affect how much these local transient fields in the QW affect the QD exciton transition linewidth inhomogeneity. We measured the linewidths of 265 MDE and 229 Q2T QDs individually, for all QDs present in 100 PL spectra studied with no selection bias, recorded with the same laser power. The linewidths of both types of QDs span from ~300 μeV to 2.5 meV, as displayed in Figure 3(a). The almost identical (no significant difference by a Student's *t*-test) statistical mean linewidths with standard errors are 892 ± 16 and 932 ± 15 μeV, for Q2T and MDE QDs respectively, with very similar distribution widths caused by the aforementioned inhomogeneous broadening.

Now we turn to examine the slow-timescale spectral diffusion in detail. For this study, we chose a pair of Q2T and MDE QDs with respective linewidths of 1.18 and 1.17 meV, emitting at 2.59 and 2.51 eV with peak intensities around 300 counts/s. When 1-s intensity-integrated spectra mapped for 100 s are displayed over a spectral window of 1 nm in Figure 3(b), the Q2T QD has a much smaller degree of slow-timescale spectral diffusion. To monitor this phenomenon more accurately, the exact energies of the QD emissions at a finer 100-ms integration time are determined with Gaussian fits. As shown in Figure 3(c), the degree of slow-timescale spectral diffusion is indeed smaller for the Q2T QD, with a standard deviation of 17.5 μeV compared to the MDE value of 70.8 μeV. While both types of *a*-plane QDs have smaller spectral diffusion compared to *c*-plane dots, the Q2T result represents an order of magnitude decrease from *c*-plane QDs, which is between 120 to 260 μeV [29]. It is also worth noting that Q2T standard deviation (17.5 μeV) is now on par with that of pure GaN QD exciton (16.4 μeV) reported recently [31], despite being attained in an InGaN system with a greater intrinsic tendency to form carrier traps and cause spectral diffusion due to indium incorporation. As discussed earlier, the reduced number of carriers trapped in many fewer sites around Q2T QDs causes a reduced degree of locally fluctuating electric field, thereby resulting in a smaller amount of QD slow-timescale spectral diffusion. This is further corroborated with a study of the spectral diffusion (1 standard deviation) of 10 Q2T and 10 MDE QDs without selection bias shown in Figure 3(d). The Q2T mean of 33.8 ± 13.9 μeV is a 65% reduction from the MDE value of 97.9 ± 51.6 μeV. The large standard deviation of MDE QDs is caused by more varying local environments and the presence of carrier traps, as opposed to the Q2T case where the formation of these traps is minimized.

Similarly, the creation of greater electron and hole wavefunction spatial overlap caused by carriers trapped around the QD can be assessed using time-resolved μ-PL. As the radiative lifetime of *a*-plane InGaN QDs are close to the width of the PMT instrument response function (IRF, ~ 100 ps), the fitting to the data has been performed with a modified Gaussian obtained from a convolution of the Gaussian-like IRF with an exponential decay function. The so determined lifetime of a Q2T QD studied in Figure 4 has been found to be 150 ± 1 ps. With this method, we perform a statistical comparison of the radiative lifetimes of 46 MDE and 36 Q2T QDs (selected without bias) displayed in the inset of Figure 4. Average lifetimes of MDE and Q2T QDs are 309 ± 25 and 173 ± 12 ps respectively, indicating that the reduction of local carrier traps has indeed increased the exciton oscillator strength by 45%. This is faster than state-of-the-art *c*-plane GaN dot-in-nanowire systems in which the extremely small size and strong spatial confinement allows lifetime to reach ~300 ps [18], as well as *m*-plane InGaN dot-on-nanowire-sidewall systems in which the residual built-in field is lower than MDE *a*-plane ones (average ~260 ps) [30]. Furthermore, despite the 50% smaller InGaN emitting energy than GaN (and noting that lifetime ∝ 1/$E_{exciton}$), Q2T *a*-plane InGaN QDs have radiative recombination lifetimes as fast as *a*-plane GaN systems [38].

## IV. SUMMARY

In summary, we have achieved critical progress with the Q2T growth method for *a*-plane InGaN QDs, significantly improved the material quality, and reduced the amount of carrier trapping sites. As a result, these improved polarized single-photon emitters have 45% shorter radiative lifetime and a 65% reduction in their slow-timescale spectral diffusion compared to MDE *a*-plane QDs, both of which are now comparable to state-of-the-art GaN systems. Coupled with greater temperature stability [39], Q2T QDs are a useful addition for the further development of electrically driven devices [40], as well as cavity-coupled systems such as porous distributed Bragg reflectors [41] with nanopillars for upcoming high density QD lasing and single-photon applications. The ease of indium control in the Q2T routine also allows more straightforward tuning of the QD emission energy during growth, and opens up possibilities for improved *c*-plane growth. However, the reliance on QW still adversely affects the purity of Q2T single-photon emission, and reduction of the InGaN epilayer thickness, whilst maintaining good sample quality, could be attempted as a solution. Optimization of the fabrication routine is also needed to increase the overall brightness, dot density, and uniformity. But as an initial development, Q2T *a*-plane InGaN/GaN QDs have already shown promising progress and provided a pathway, which with further development, could lead towards both integrated quantum photonics and the unveiling of unknown fundamental photophysics.

## ACKNOWLEDGEMENTS

This research was supported by the U.K. Engineering and Physical Sciences Research Council (EPSRC) Grants No. EP/M012379/1 and EP/M011682/1. T.W. is grateful for the award of a National Science Scholarship (NSS) as Ph.D. and postdoctoral funding by the Singapore Agency for Science, Technology and Research (A*STAR). C.C.K. is grateful for the support provided by an EPSRC scholarship, a Clarendon Scholarship and a Mary Frances and Philip Wagley Graduate Scholarship. R.A.O. is grateful to the Royal Academy of Engineering and the Leverhulme Trust for a Senior Research Fellowshi

All data presented in Figures 1–4 are available via *url to be added after review*.

Figures

Figure 1

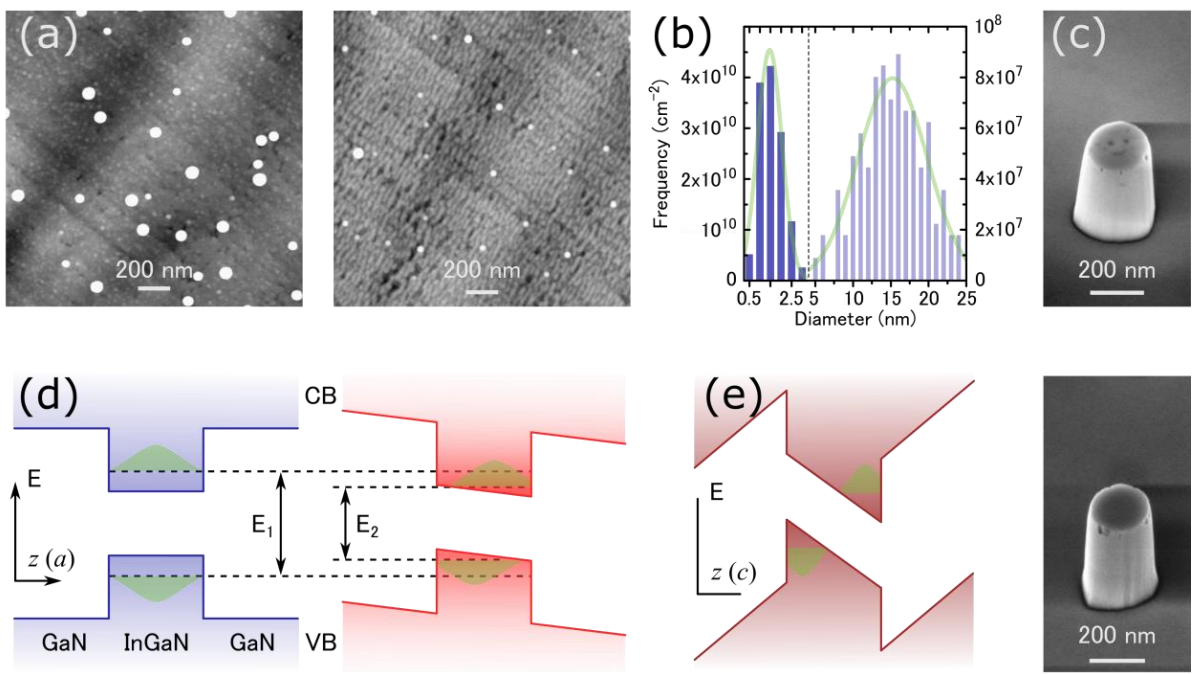

Figure 2

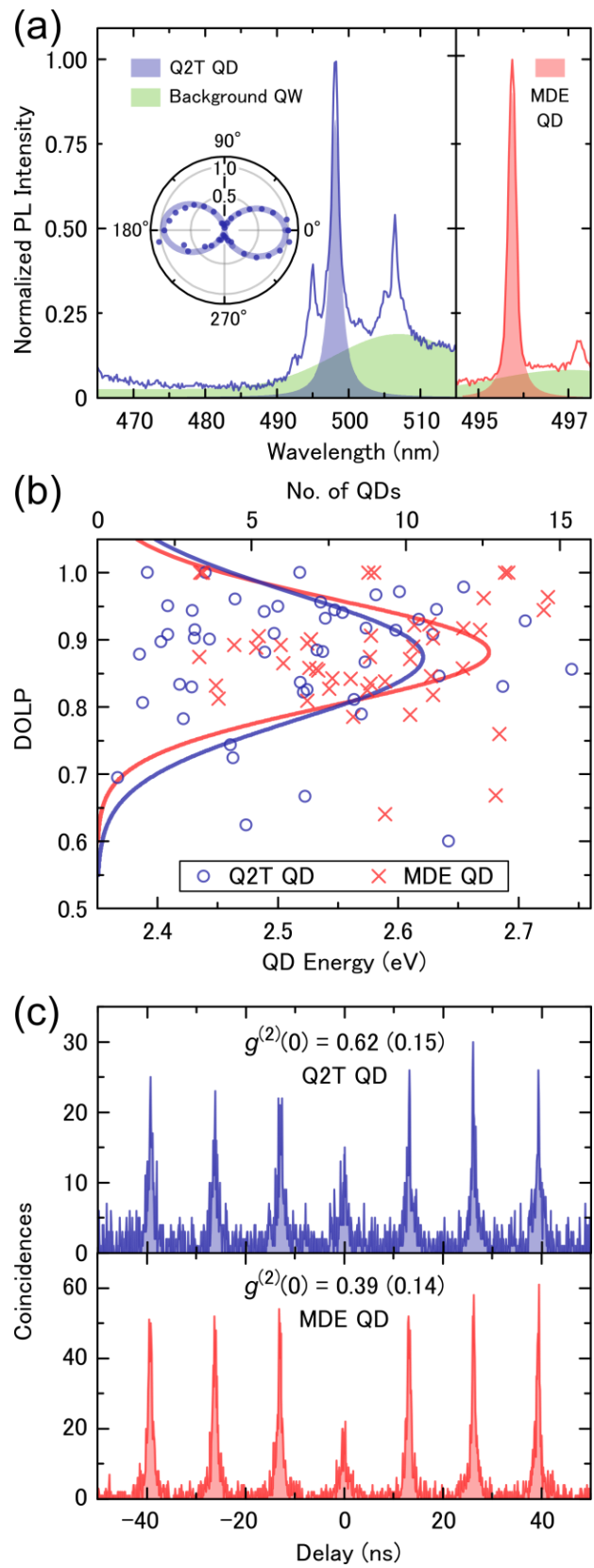

Figure 3

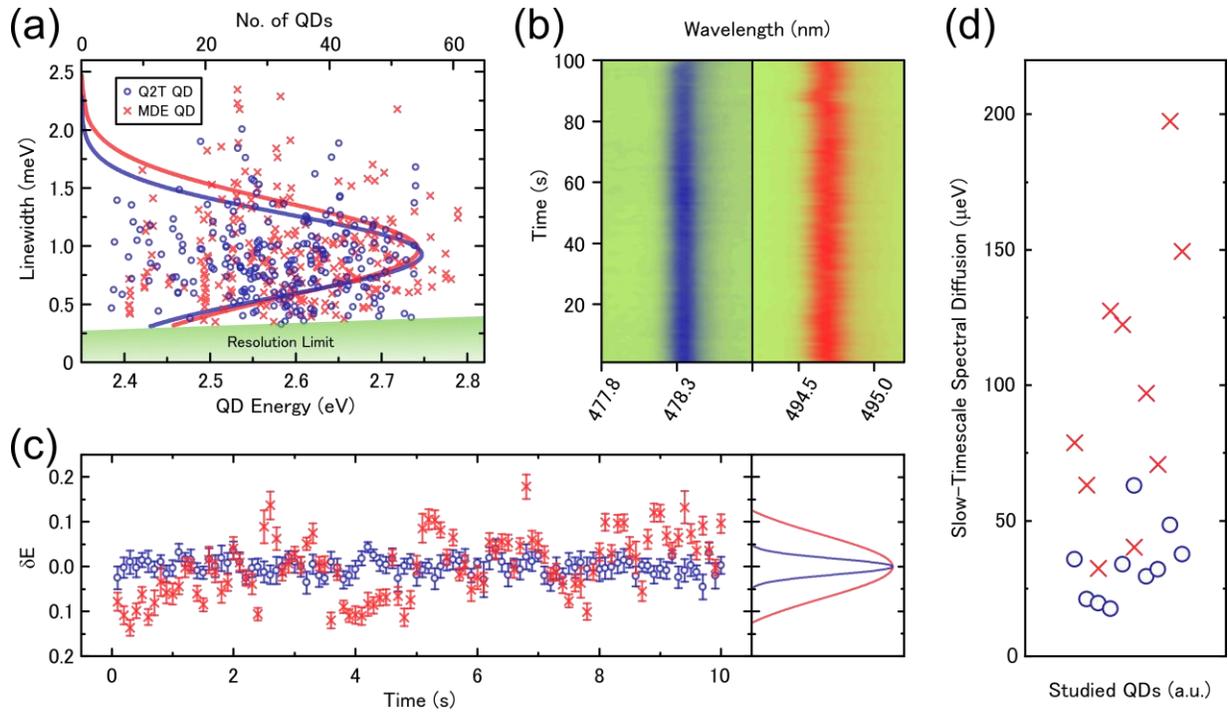

Figure 4 *(1-column, ideally page 4/5 left/right)*

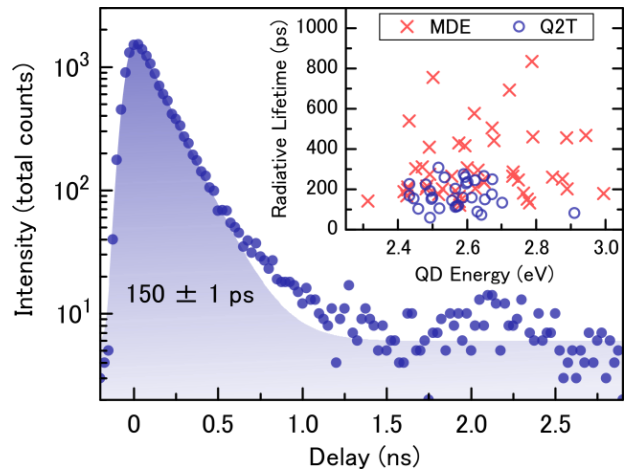

Figure Captions

FIG. 1. (a) AFM images of uncapped *a*-plane InGaN QDs grown by the Q2T (left) and MDE (right) methods. Bright spots are indicative of metallic droplets thought to re-react during capping for QD formation. (b) Statistics of the diameters of uncapped Q2T QDs, exhibiting a bimodal distribution. Both components of the distribution have been fitted with Normal distributions with mean values at 1.5 and 15 nm. (c) SEM images of postprocessed nanopillars for Q2T (left) and MDE (right) samples for enhanced photon extraction efficiencies. (d) Illustration of the ideal non-polar InGaN energy profile with no external fields present (left) and with a small external field caused by carriers trapped in the vicinity of the QD (right). The latter causes a small spatial separation of electron and hole wavefunction as well as a change in emission energy. (e) Contrast to the energy profile of conventional polar InGaN, where the slope of energy levels (not drawn to scale) and the resultant wavefunction separation and emission energy change are much more drastic.

FIG. 2. (a) µ-PL spectra of representative Q2T (Q2T) and MDE (right) *a*-plane InGaN/GaN QDs. The studied Q2T QD at ~ 498 nm, together with several other weaker QDs at different wavelengths, spectrally overlaps with the underlying InGaN QW emission, which is weaker in the MDE case. Inset: Peak intensities at varying polarizer angles, fitted with a sinusoidal function, demonstrating the linearly polarized nature of the emitted photons. (b) Statistical comparison of the DOLP of 50 Q2T and 50 MDE QDs. The DOLP distributions have been fitted with Gaussian functions and demonstrated no change in the intrinsic polarization properties between two growth methods. (c) Photon antibunching measurement for the studied Q2T (top) and MDE (bottom) QDs. Raw $g^{(2)}(0)$ are displayed together with parenthesized background-corrected values. Together with (a) and (b), these results confirm the ability of Q2T QDs to act as polarization-controlled single-photon sources.

FIG. 3. (a) Exciton transition linewidth (fast-timescale spectral diffusion) comparison between 229 Q2T and 265 MDE QDs, without selection bias. The Gaussian distribution shows very similar average linewidths for both types of QDs. With both methods, a number of QDs with linewidths close to the spectral resolution are found. (b) PL mapping at 1-s step over a 100-s period for two representative Q2T (left) and MDE (right) QDs with similar emission energy and near identical linewidth, displayed over a 1-nm window. A larger degree of spectral wandering (slow-timescale spectral diffusion) is present in the studied MDE QD. (c) Measurement of emission energy drift for a 10-s period at a 100-ms integration time for QDs studied in (b). The energy distributions have been fitted with Gaussian functions. (d) Comparison of the emission energy standard deviation between 10 Q2T and 10 MDE QDs, demonstrating a reduction in both average magnitude and spread of the slow-timescale spectral diffusion of Q2T QDs.

FIG. 4. Radiative lifetime of a typical *a*-plane QD grown by the Q2T method. A convolution of a Gaussian and an exponential decay has been used to fit the data. Inset: statistical comparison of the radiative lifetimes of 33 Q2T and 46 MDE QDs. A much smaller mean and standard deviation are found for Q2T QDs.